\documentclass[reprint,superscriptaddress,amsmath,amssymb,pra ]{revtex4-2}

\usepackage{graphicx}
\usepackage[utf8]{inputenc}

\begin{document}

\title{Experimental decoherence in molecule interferometry}
\author{Markus Arndt}
\affiliation{University of Vienna, Faculty of Physics, Boltzmanngasse 5, 1090 Vienna, Austria.}
\email{markus.arndt@univie.ac.at}
\homepage{www.quantumnano.at}
\author{Stefan Gerlich}
\affiliation{University of Vienna, Faculty of Physics, Boltzmanngasse 5, 1090 Vienna, Austria.}
\author{Klaus Hornberger}
\affiliation{University of Duisburg-Essen, Faculty of Physics, Lotharstra{\ss}e 1-12, 47048 Duisburg, Germany.}

\date{\today} 

\begin{abstract}
Complex molecules are intriguing objects at the interface between  quantum and classical phenomena. Compared to the electrons, neutrons, or atoms studied in earlier matter-wave experiments, they feature a much more complicated internal structure,
but can still behave as quantum objects in their center-of-mass motion. 
Molecules may involve a large number of vibrational modes and highly excited rotational states, 
they can emit thermal photons, electrons, or even atoms, and they exhibit large  cross sections for collisional interactions with residual background gases. This makes them ideal candidates for decoherence experiments which we review in this contribution.
\end{abstract}

\maketitle

\section{\label{EDsec:Intro}   Emergence of Classical Behavior in Quantum Experiments}

This contribution is written in honor of H.~Dieter Zeh, a pioneer in  developing our understanding of the transition from  quantum behavior to the appearance of classicality \cite{Zeh1970}. Starting with his  work with E.~Joos \cite{Joos1985}, decoherence theory has evolved from a stance on the foundations of quantum physics 
to a practical guide of what must be done and what has to be averted
if one seeks to observe quantum phenomena in increasingly complex and macroscopic systems.
Indeed, current advances in quantum technologies  such as matter-wave interferometry, quantum-assisted sensing and metrology, quantum simulation, and quantum computing owe their success to no small degree
to a better understanding and control of decoherence phenomena. 

In a nutshell, decoherence theory focuses on how a quantum system affects its environment. Unavoidable interactions  correlate the quantum state of the system with its environment in a way that quantum information gets disseminated into the uncontrolled surroundings. Since this information is in principle available to an outside observer, the system  behaves effectively as if it had been subjected to a measurement device with the outcome being discarded. The associated loss of coherence can be much faster than all other time scales in the experiment, and typically exceeds by many orders of magnitude other environmentally induced rates, such as friction or thermalization. Decoherence theory is by now a well-accepted field of quantum physics; introductions to the general concept and ideas can be found in \cite{Joos2003,Zurek2003a,Hornberger2008,schlosshauer2019quantum}.

Decoherence is often put forward as a solution to the measurement problem of quantum mechanics. However, the theory involves no collapse of a wave function. It presupposes  unitary quantum dynamics and the superposition principle to hold not only for the  quantum system but also for its environment and, ultimately the ``universe''.
Decoherence theory therefore cannot explain why the pointer of a macroscopic measurement apparatus ends up at a specific position of the dial instead of any another, that would be equally accessible from  the initial state. But it can answer  why no such  pointer is ever encountered in a superposition of positions, i.e.\ it can predict  when and how the environment selects the basis of pointer states \cite{Zurek1981,Zurek1982}.
Decoherence theory has found sympathy among those who 
think that quantum physics can be explained without invoking a measurement postulate \cite{Zurekdarwinism}, e.g.\ by adhering to a many-worlds interpretation.
While no experiment  so far has been able to  discriminate between interpretations of quantum mechanics, work in the lab can clearly corroborate the validity of decoherence theory and emphasize its importance for our understanding of modern quantum technology.

From an experimental point of view, genuine decoherence is often hard to distinguish from effects of \emph{phase averaging} that may lead to a similar loss of coherence in the observed quantum system. In phase averaging an interference phenomenon is washed out because the quantum evolution depends on noisy classical parameters that are uncontrolled in practice, but could be measured in principle. For instance, the fringes in matter-wave interferometry may get blurred due to vibrations or fluctuating external electromagnetic fields. The interference contrast could then  be improved by monitoring the grating accelerations or field strengths and by correlating these signals with the detector recordings. 
This is to be contrasted with genuine decoherence, where the quantum nature of the environmental degrees of freedom and their entanglement with the quantum system rule out regaining coherence,  because Hilbert space and state manipulation grow exponentially with the number of degrees of freedom. In the following we will therefore make a clear distinction between the phenomena of decoherence and phase averaging.

\section{Experimental decoherence in macromolecule interferometry}
The \emph{decoherence program} of H.~Dieter Zeh and his coworkers has inspired a plethora of different  experimental studies. While in the 90's of the last century one could still occasionally hear the question "Do you really believe in decoherence theory?", this framework has meanwhile become accepted and respected as what it is: the consequent application of unitary quantum mechanics to complex systems. Decoherence has also gained importance in the rapidly growing field of quantum information research, where it is the ultimate challenge in quantum computing~\cite{Nielsen2010} and quantum simulation~\cite{Tacchino2019}. The progressive coupling of Schr\"odinger-kittens to the environment has also been experimentally demonstrated on the level of a single field mode, by probing high-finesse microwave cavities with circular Rydberg atoms~\cite{Brune1996}. Many related studies have been performed in a large variety of systems. 

Here, we focus on applications in macromolecule interferometry as performed at the University of Vienna. While the concepts are universal, we illustrate them in the specific context of the interferometers that we have built throughout the last two decades: from far-field diffraction  (FFD) of hot molecules \cite{Arndt1999,Juffmann2012,Brand2015}, via the first  Talbot-Lau interferometer (TLI) with molecules  \cite{Brezger2002}, Kapitza-Dirac-Talbot-Lau interferometry (KDTLI) for highly polarizable molecules \cite{Gerlich2007},  Talbot-Lau lithography with fullerenes (TLL) \cite{Juffmann2009}, Optical Time-Domain Matter-Wave interferometry (OTIMA) with organic clusters \cite{Haslinger2013} and antibiotic polypeptides~\cite{Shayeghi2020} to the most recent addition in the series, the Long-Baseline Universal Matter-Wave interferometer (LUMI) \cite{Fein2019}. The latter sets the current complexity record for matter-waves,  with molecules exceeding the mass of $ 25 $\,kDa. We refer to existing literature on how to prepare the sources, detectors, and interferometers \cite{Hornberger2012,Juffmann2013,Arndt2014} and also for novel concepts in nanoparticle interferometry \cite{Arndt2014,Stickler2018}. 

\subsection{Coherence in the molecular beam} 
\label{EDsec:coherence}
The formation of a molecular beam requires a volatilization mechanism. For many stable molecules up to $m\simeq 1000\,$Da this is most simply realized by sublimating or evaporating them in a Knudsen cell. For large organic molecules or complex peptides this is done most efficiently using nanosecond or even femtosecond laser desorption\cite{Schaetti2018}. In some cases, the molecules reach temperatures up to 1000\,K. How can we understand that even at such high temperatures coherence can be prepared and maintained?   
The longitudinal velocity of a thermal beam follows a  distribution $f(v)$, which in practical applications is often a Maxwell-Boltzmann distribution shifted by $v_0$ towards higher velocities, $f(v)\propto v^2 \exp\left [-m(v-v_0)^2)/2 k_B T \right ]$. This is due to collisions in the source. 

Particle beams with a high temperature and mass give rise to  a distribution of small de Broglie wavelengths, e.g. from $ 5\,$pm for thermal fullerenes \cite{Nairz2003} down to $50$\,fm for hot oligoporphyrin derivatives at $m=25$\,kDa and $v=300$\,m/s \cite{Fein2019}. While these wavelengths are extraordinarily small in comparison to those encountered in atom interferometry, and even though they can be up to $10^5$ smaller than the diameter of the molecule itself, these are no fundamental constraints to matter-wave coherence. 
This is because  the effective transverse motional temperature is usually orders of magnitude lower than the longitudinal temperature found in an effusive source: tight collimation to about   
$\theta_\mathrm{coll}\simeq 10\,\mu$rad in far-field diffraction can select a transverse molecular temperature down to $10\,$nK, in exchange for a dramatically reduced beam flux. In near-field interferometry the collimation requirement is typically one hundred times less strict, associated with a  transverse temperature on the order of $100\,\mu$K.

In molecular beams with a broad longitudinal velocity distribution 
the interference pattern is still an average of the fringes associated with fixed longitudinal velocities, as described by the correspondingly short longitudinal coherence length $L_c \propto \lambda_\mathrm{dB} v/\Delta v$. Also the high internal temperature, which can often be assumed equal to the motional temperature in effusive cells, is a specific feature of many  experiments in molecule interferometry.  

\subsection{\label{EDsec:MolecularDephasing} Phase averaging and dispersion }
Since coherence signifies the persistence of constant phase relations between different parts of the matter wave across space or time, one may be  tempted to denote anything destroying this relation as  decoherence. It is, however, often useful to be more precise and to distinguish phase averaging and decoherence. The former can be caused by a variety of fluctuating parameters of the experiment---changes in the source position, vibrations of the experimental setup, or random noise in electric or magnetic fields. In contrast to that, we reserve the denotation \textit{decoherence} for processes that lead to quantum correlations between the interfering particle and an (internal or external) environment.

\subsubsection{Phase averaging by apparatus noise\label{EDsec:ApparatusDephasing}}
Noise due to vibrations in the interferometer is at the same time the most important and the most trivial reason for loss of coherence. The phase difference associated with neighboring paths through a 3-grating interferometer, $\Delta \varphi = (x_1 -2x_2 +x_3)\cdot 2\pi/d$, is determined by the positions $x_i$  and the  period $d$ of all gratings. This shows that controlling all of these positions on the nanometer level may be required to maintain a stable interferometer phase. While this constitutes a substantial experimental challenge, grating vibrations can today be well controlled using passive or active isolation systems, even in demanding cases~\cite{Peters1999,Fein2019}. 

Some of the best current atom interferometers aim at a relative precision in accelerometry of better than $10^{-10}$ to $10^{-12}$. While it is difficult to eliminate all vibrations, precise phase readings can still be achieved by correlating complementary  interferometer output ports. Even if each of them exhibits random phase noise complementary ports remain highly correlated. This way a differential readout can suppress common mode noise to an impressive level~\cite{Parker2018,Asenbaum2020}.  

\subsubsection{Van der Waals dispersion \label{EDsec:vdWDephasing}}
Because of its conceptual and technological simplicity, far-field diffraction of the fullerene C$_{60}$ \cite{Arndt1999} was the first experiment to demonstrate the matter-wave nature of hot large molecules. The molecular volume, structure, and internal excitation contributes already several interesting aspects to our discussion on phase averaging and decoherence. Naively speaking,  \textit{de Broglie} interference should be characterized only  by the value of the de Broglie wavelength $\lambda_\mathrm{dB}=h/mv$, determined by Planck's quantum of action $h$, the particle's mass $m$ and velocity $v$. However, in practice the interaction with the gratings depends on several molecular properties. 

Fullerenes are composed of 60 covalently bound carbon atoms, arranged in the shape of a miniaturized soccer ball of $0.7$\,nm inter-nuclear diameter. Because of their size and electron delocalization across the bonds, the static electric polarizability of fullerenes is a sizeable
$\alpha_\mathrm{stat} = 4\pi \varepsilon_0 \times 89$\,\AA$^3$ \cite{Berninger2007,Fein2020a}.    
Even in absence of a static molecular dipole moment its quantum fluctuations  will induce image dipoles in the grating, which are always oriented to attract the particle to the grating wall. 
This Casimir-Polder potential V$_\mathrm{CP}$ \cite{Casimir1948} is often distinguished from the simpler van der Waals approximation $V_\mathrm{vdW}(r)=-C_3/r^3$, valid at close distances $r$, and the asymptotic far-distance behavior $V_\mathrm{CP}(r)=-C_4/r^4$. The constants $C_3$ and $C_4$ depend on the particle polarizability and the dielectric properties of the surface. The  Casimir-Polder  interaction is conservative and does not introduce any decoherence or phase averaging, even though it is determined by the the internal particle structure. 

Due to the presence of this dispersion force the matter-wave  will in general  acquire a sizeable phase shift
$\Delta \varphi = \int dz V_\mathrm{CP} /\hbar v$
when passing the grating structure.
It depends strongly on both the particle's velocity and its distance to the grating. 
In {far-field diffraction} of atoms or molecules, this effect may already substantially modify the diffracted intensity distribution, favouring higher diffraction orders over lower ones \cite{Grisenti2000,Lonij2010,Nairz2003,Brand2015}. In {near-field interferometry}, the strong dispersion can become influential \cite{Brezger2002} and even detrimental: already for fullerenes travelling at 200 m/s across a Talbot-Lau interferometer with a grating period of $d=266$\,nm, the acceptable velocity band can shrink to $\Delta v/v \simeq 0.5 \%$. Even though all these velocity-dependent phase shifts are well understood, a finite longitudinal velocity distribution will quickly superimpose too many modified interference patterns, such that  the fringe visibility gets lost. 

The solution to this challenge is to either  develop molecular sources with sufficiently good velocity selection and coherence length--a mayor aim for several years to come--or to replace the nanomechanical diffraction gratings by standing waves of light. The phase shift in optical gratings is still dispersive but the distance dependence in a sinusoidal structure is much more forgiving than the $r^{-3}$-dependence of the van der Waals potential. Theory and many experiments have proven that light gratings facilitate high contrast matter-wave interference, over large velocity bands and even for highly polarizable molecules \cite{Gerlich2007}.

\subsubsection{Rotational effects\label{EDsec:RotationalDephasing}}

In contrast to atoms, most molecules cannot be considered spherical. They are revolving as aspherical rotors, implying that any orientation-dependence of the molecule-wall interaction may have an impact on the interference pattern. Indeed, large polarizability anisotropies can be found in molecules such as carotenes, phthalocyanines or porphyrins.
Since massive molecules require low velocities to stay within acceptable bounds for their de Broglie wavelengths and since the polarizability typically increases with molecular mass, 
the impact of anisotropies in large molecules is further increased due to the greater grating interaction time.
In addition, all molecules without perfect inversion symmetry, i.e.\ the vast majority of molecules, exhibit a permanent electric dipole moment. 

Most  molecule-wall interactions therefore depend on the orientation state of the particle. The latter can be usually characterized only by a rotational temperature, since neither the initial orientation nor the particle's angular momentum can be controlled in any currently existing or foreseeable molecular beam source. Even tiny local charge imbalances inside the grating slit will then diffract the molecule differently for different orientations with respect to the slit \cite{Stickler2015}. This expectation is consistent with the observation that polar molecules have never led to sizable fringe contrast in either far-field diffraction at nanogratings \cite{Knobloch2017} or near-field interferometry at larger gold gratings \cite{Hackermueller2003}. 
However, as suggested by theory and confirmed by experiment, 
high-contrast interference of polar molecules is still possible upon diffraction at optical phase gratings \cite{Hornberger2009} or photo-depletion gratings \cite{Shayeghi2020}.

\subsubsection{The role of molecular vibrations \label{EDsec:VibrationalDephasing}}
Molecules are no static entities; apart from rotating they may vibrate or even undergo structural state changes in free flight. Typical time scales of these processes for molecules with about the size of C$_{60}$ or beta-carotene are 100\,fs to 1000\,fs for vibrations, 100\,ps for rotations and 1\,ns for structural changes.  It is remarkable that the electric dipole moment may  
fluctuate by as much as 300\% in hot functionalized azobenzenes \cite{Gring2010} or $\alpha$-tocopherol \cite{Mairhofer2017}, and that we still observe high-contrast matter-wave interference with these molecules, even in the presence of external electromagnetic fields and field gradients. 

This can be explained by noting that on the fast time scale of the dipole fluctuations
the external field varies adiabatically slow. In consequence, if an electric field $\mathbf{E}$ is applied inside the interferometer all that happens is that the interference pattern gets shifted. The associated force $\chi_{\rm tot} (\mathbf{E}\nabla) \mathbf{E}$ is determined by a time-averaged scalar quantity, the electric susceptibility $\chi_\mathrm{tot}=\alpha_\mathrm{stat}+\langle \mathbf{d}\cdot\mathbf{d} \rangle/3 kT$.
It involves not only  the static polarizability $\alpha_\mathrm{stat}$ but also the temporal average of the thermally fluctuating dipole moment $\mathbf{d}$, with $T$ the vibrational temperature of the molecule.
Figuratively speaking, one may thus say that the vibrational effect on the interference fringes allows one taking a look into the molecule and its thermally averaged nanosecond dynamics. This holds even though nature forbids us to know which path the molecule took during its flight through the interferometer.

\subsubsection{Inertial forces \label{EDsec:InertialDephasing}}
If a matter-wave interferometer consisting of  three equal gratings is accelerated it accumulates a phase  $\varphi = k_\mathrm{eff} a T^2$. Here $k_\mathrm{eff}$ is the effective reciprocal lattice vector of the diffraction gratings, $a$ the transverse acceleration and $T$ the transit time between two subsequent gratings. Gravity would cause an acceleration of $a\simeq 9.81\,$m/s$^2$, while a rotation of the interferometer around the slit direction at angular frequency $\Omega$ will cause a Coriolis acceleration $a=2v\Omega $ for particles of longitudinal velocity $v$. 

To date, no molecule interferometer has been optimized towards sensitivity to these inertial forces and it would be difficult to compete with atom interferometry because of the prevalent phase-space density and signal-to-noise ratio in current molecular beam sources. However, all advanced molecular de Broglie experiments are significantly influenced by both gravity and Earth's rotation. While an overall phase and fringe shift would not matter in many experiments, a finite velocity spread $\Delta v$ entails a finite   phase distribution $\Delta \varphi= 2 k_\mathrm{eff} a T^2  (\Delta v/v)$. To observe high-contrast fringes in the interference pattern  we require $\Delta \varphi\ll 2\pi$. In consequence, molecule and high-mass matter-wave interferometers need to either compensate both gravity and the Coriolis force by an appropriate choice of the interferometer alignment \cite{Fein2020a}, or the  velocity spreads must be narrowed down sufficiently by beam cooling or velocity selection.

\begin{figure}
   \centering
   \includegraphics[width=0.5\textwidth]{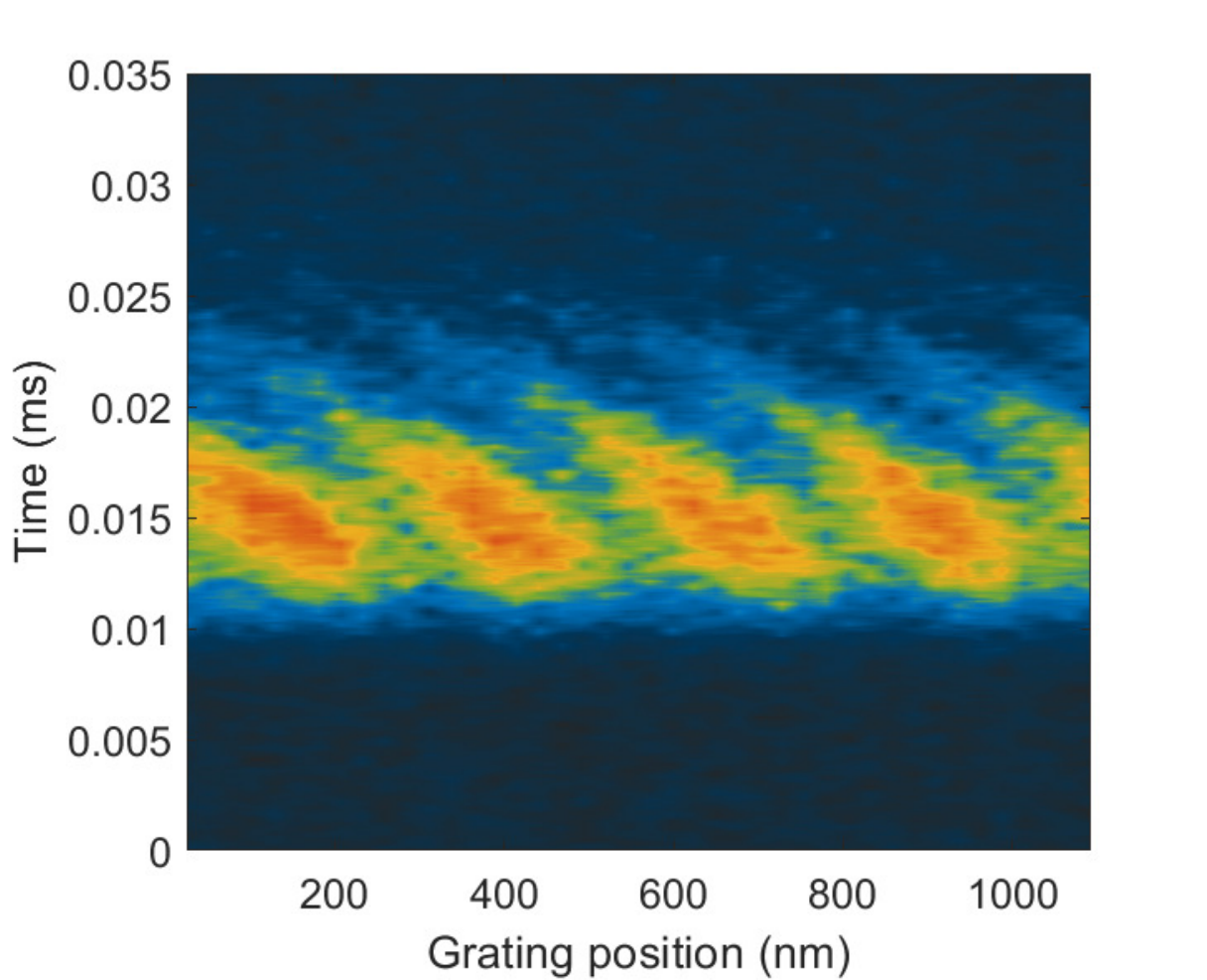}
    \caption{Matter-wave fringes in fullerene interferometry for different flight times of C$_{60}$ molecules through the LUMI interferometer (with grating period 266\,nm). The Coriolis force entails a velocity dependence of the phase of the interference fringes. The phase difference between fast and slow molecules can easily exceed $2\pi$. (Reproduced from Ref.~\cite{Fein2020a})}\label{EDfig:Coriolis}
\end{figure}

Figure \ref{EDfig:Coriolis} shows the effect of the Coriolis force on fullerene interference fringes in the LUMI interferometer. Averaging over different velocities here corresponds to a vertical sum over each column of that image, which would reduce the contrast substantially.

\subsection{Decoherence due to internal states}

\subsubsection{The fallacy of the internal clock}
Atoms turn into optical or radio-frequency clocks when we prepare and probe superpositions of their electronic or hyperfine  states. In analogy, complex molecules may 
be viewed as rotational or vibrational clocks. The analogy to a clock becomes intuitive when we consider rotating polar molecules, where the dipole moment acts as the hand of the clock.

Consider for instance  the antibiotic ciprofloxacin, C$_{17}$H$_{18}$FN$_3$O$_3$, which carries many different clocks. It is composed of 41 atoms with  a total molecular weight of 331.3\,Da. It has 123 different degrees of freedom, shared among three directions of translation, highly excited rotations around three randomly oriented axes and 117 vibrational modes. In its thermal ground state, the molecule has an electric dipole moment of 2.7 Debye, which may be regarded as a hand of a clock, rotating at a frequency of the order of $10^{10}\,$Hz when heated to about 500\,K. In addition every vibrational normal mode is populated and `counts time' on its own.

 When the molecule is diffracted at a thin standing light wave, first-order far-field diffraction requires two neighboring paths through the grating towards the detector to differ by a single de Broglie wavelength. Naively, one might wonder whether this path-length difference is associated with different arrival times that can be read off from the clock, which would reveal which-path information and cause decoherence. However, interference and detection of each molecule takes place at a well-defined time, no matter whether it had existed as a superposition of different wavelets  inside the interferometer.
Since external and internal degrees of freedom are decoupled, the same molecule arriving at the same detector at the same time must be in the same internal state and the internal clock remains in phase. 
The prize to pay is coherence: the longitudinal coherence of the  molecular beam must be sufficiently large for the high velocity components in the longer arm to  catch-up with low velocity components in the shorter arm of the interferometer. Like for point particles without internal clocks, the longitudinal coherence length $L_c$ discussed in Sect.~\ref{EDsec:coherence} must  exceed $n \lambda_\mathrm{dB}$  for $n^\mathrm{th}$-order far-field diffraction to be observable \cite{Arndt2005}. 

\subsubsection{Molecular internal clocks in the gravitational field}
\label{EDsec:clocksingravitationalfield}
The issue of the internal clock has  been recently revisited in the context of the gravitational time dilation predicted by general relativity: What happens if we prepare an atom in a quantum superposition of two internal clock states and propagate this ticking clock in a superposition along two vertically separated arms of an interferometer? Given that the red shift in the gravitational field of the Earth amounts to $\Delta \nu/\nu \simeq 10^{-16}$ per meter height difference, how long can that superposition be maintained before the proper times in the upper and the lower arm differ sufficiently to prevent any interference due to the  which-path information provided by the clock? 

As analyzed in Ref.~\cite{Zych2011} the gravitationally induced dephasing between internal and external degrees of freedom should become observable in near-future rare-earth atom interferometers featuring path separations throughout seconds and on the meter scale. However, unlike genuine decoherence, this loss of coherence is expected to be fully reversible: if the spatial superposition is maintained for sufficiently long, the internal clock states will oscillate between being in and out of phase, periodically admitting or preventing matter-wave interference. 

One may extend this idea to high-mass molecules as typically used in KDTLI or LUMI interferometry. These molecules rotate and vibrate in typically 6000  modes, most of them ticking at different frequencies. The above-mentioned interference revivals should then be suppressed or strongly delayed, thus providing a universal path to classical behavior whenever the rotational or vibrational clock is red-shifted in the gravitational field \cite{Pikovski2015}. 
However, the frequencies of vibrational modes are about three orders of magnitude smaller than the transition frequencies in optical clocks and  existing  beam splitters allow splitting the molecular beam over microns rather than on the meter scale.
To observe this type of red-shift decoherence in large molecules one still needs to bridge more than eight orders of magnitude in the product of beam separation and coherence time. Increasing the mass is less favorable here, as the required coherence time scales with the particle mass as $1/\sqrt{M}$.

\subsection{Decoherence due to the environment}
Decoherence due to momentum transfer with environmental degrees of freedom is by far the most direct and most relevant mechanism in macromolecular quantum optics. Already 14 years before the first fullerene diffraction was realized in Vienna\,\cite{Arndt1999},  decoherence of high-mass superpositions was studied by E.~Joos and H.~Dieter Zeh \cite{Joos1985}. They considered several mechanisms of which collisional and thermal decoherence were seen as the most dominant processes. 

In the decoherence literature one often finds comparisons of different decoherence rates for a delocalized quantum particle, that are obtained by linearizing its interaction with the environment. These rates grow rapidly above all bounds as the delocalization length is increased. For the usual scattering-type interactions this is clearly unphysical, an artifact of the linearization, since the decoherence rate cannot exceed the total scattering rate.
In this context, the historical note may be in order that in the early days of decoherence studies some theory colleagues declared that fullerene diffraction will never be observed---based on a linearized description of gas scattering.
This only illustrates that all theory must be put into context and that one must be aware of the  underlying assumptions. 

\begin{figure}
    \centering
    \includegraphics[width=0.45\textwidth]{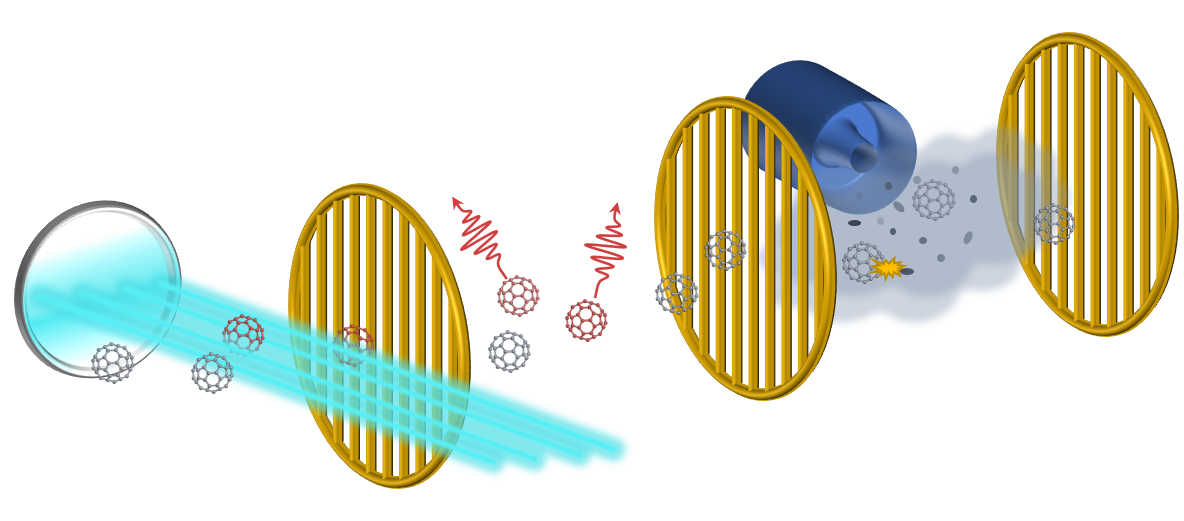}
      \caption{Molecules traversing a Talbot-Lau interferometer may lose coherence by emission of thermal radiation or collisions with residual gas in the vacuum chamber. Both processes have been studied in fullerene interferometry~\cite{Hornberger2003,Hackermueller2004}. The internal temperature can be adjusted by laser heating before the interferometer and calibrated by the enhanced electron emission close to the heating stage. The number of collisions can be set by additional gas admitted to the chamber, and controlled to a few percent with an ionizing pressure gauge.} \label{EDfig:Decoherence_Sketch} 
\end{figure}

\subsubsection{Collisional decoherence}
A first systematic study of collisional decoherence was realized \cite{Hornberger2003a} and theoretically analyzed \cite{Hornberger2003} in Talbot-Lau interferometry of fullerenes in Vienna \cite{Brezger2002} (see Fig.~\ref{EDfig:Decoherence_Sketch}). An array of $475\,$nm wide slits with a period of $991\,$nm  had been etched in to three gold gratings of $500\,$nm thickness. They were positioned in a distance of $440\,$mm from each other to form a near-field matter-wave interferometer with noticeable but still tolerable influence of Casimir-Polder forces in the molecule-grating interaction. High-contrast interference of about $30\%$ fringe visibility could be routinely observed and this quantity was monitored as a function of the background pressure for different collision gases (decohering agents). 

\begin{figure}
    \centering
    \includegraphics[width=0.5\textwidth, trim=10 20 15 30,clip]{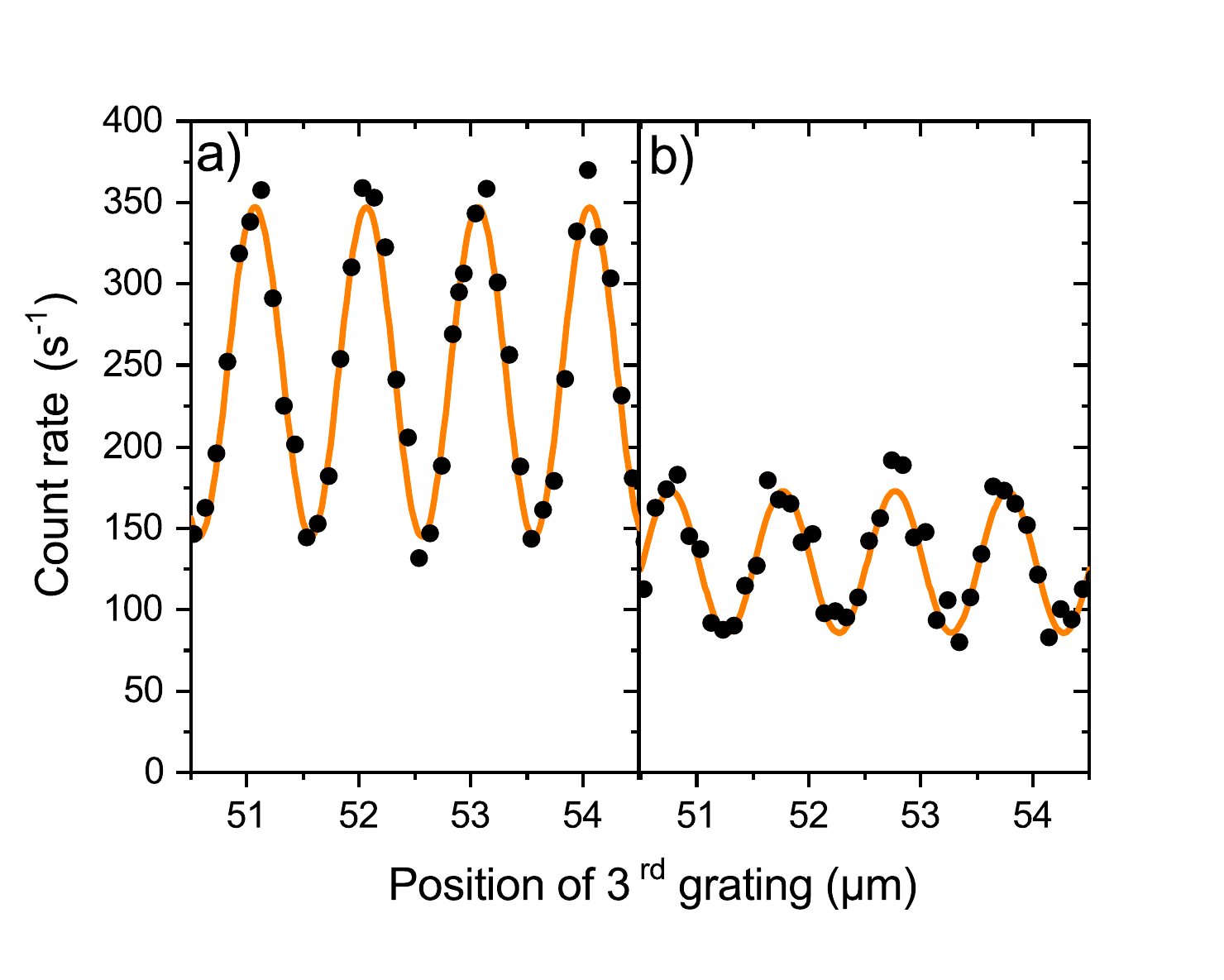}
    \caption{\textbf{a)} Typical interferogram of C$_{70}$ molecules in a TLI at a residual gas pressure of $ 3 \cdot 10^{-8}$mbar. The observed visibility of 42\% is in good agreement with theory. \textbf{b)} Already a slight increase of the background gas pressure to $ 5 \cdot 10^{-7}$mbar due to the addition of argon gas leads to a significant drop of the interference contrast to 34\%, and to a loss of counts as a certain fraction of the molecules are scattered outside the detection area. (Reproduced from Ref.~\cite{Hackermueller2003a})
    \label{EDfig:Collisions_Interferograms}
    } 
\end{figure}

Linearly raising the base pressure in the chamber from $10^{-7}\,$mbar to $10^{-6} \,$mbar reduced the contrast exponentially down to the background noise, see Fig.~\ref{EDfig:Collisions_Interferograms}. 
Also the mean count rate decreased somewhat, in agreement the Lambert-Beer law, but the transmission remained sufficiently high to allow for reliable measurements. While head-on collisions would kick the molecules to beyond the detector range, small angle collisions were of particular relevance. Even a deflection of the fullerene by 1\,$\mu$rad could reduce the interference fringe contrast, since it would shift the fringe  between a  maximum and a minimum. The cross section for such small-angle collisions, determined by the long-range Casimir-Polder interaction between two molecules, can exceed the geometrical cross section by almost two orders of magnitude, as was the case for interactions between the interfering fullerenes and nitrogen molecules~\cite{Hornberger2003}.

\begin{figure}
    \centering
    \includegraphics[width=0.5\textwidth, trim=55 10 55 55,clip]{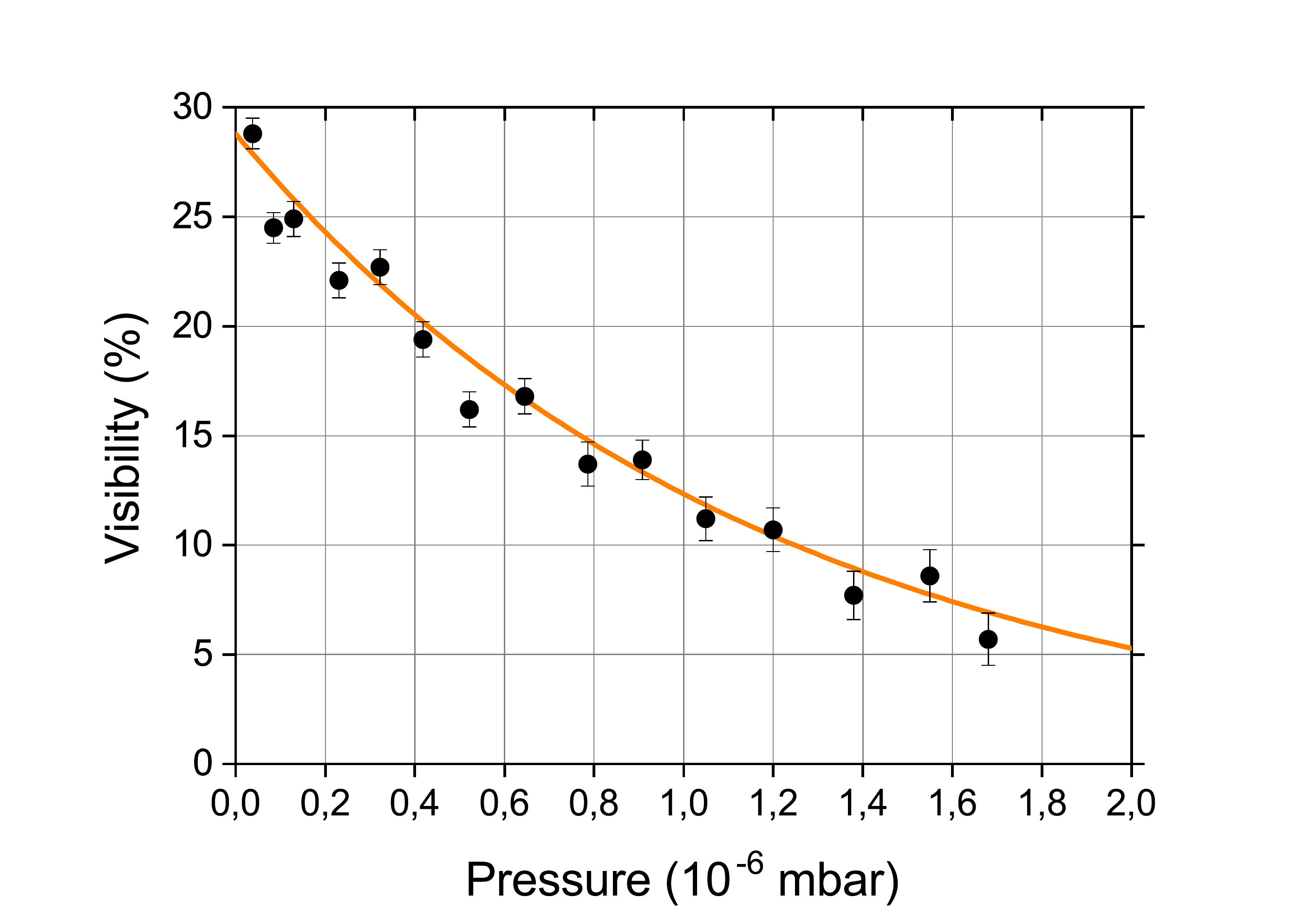}
    \label{EDfig:Collisions_Visibility}
    \caption{Reduction of the interference visibility of C$_{70}$ molecules in a TLI due to collisions with the background gas (argon) as a function of gas pressure. The data is in excellent agreement with the expected exponential loss of contrast (solid line). (Reproduced from Ref.~\cite{Hackermueller2003a})} 
\end{figure}

\subsubsection{Emission of thermal radiation by the molecule}
Molecules and nanoparticles differ from atoms, neutrons, or electrons in that they can be regarded as small thermal bodies. Given the  number $N$ of atoms, covalently bound into a single entity, they provide a reservoir of $3N-6$ modes to store vibrational energy. 

C$_{60}$ fullerenes are therefore loaded with about 6--7\,eV of vibrational energy in 174 modes when they leave an effusive source at 900\,K, and it is possible to  heat them temporarily to microcanonical temperatures beyond 3000\,K. 
Three competing processes can then be observed, which justify the view that  a single fullerene may already be regarded as a lump of classical matter: hot fullerenes emit radiation in a black-body-like spectrum \cite{Hansen1997}, they evaporate carbon dimers \cite{Dresselhaus1998}, and they undergo thermionic emission \cite{Campbell1991}, that is they emit electrons like a tungsten wire in a light bulb. Either of these processes can result in substantial decoherence if it occurs while the molecules are delocalized in a matter-wave experiment. 

\begin{figure}
    \centering
    \includegraphics[width=0.50\textwidth, trim=80 40 110 100,clip]{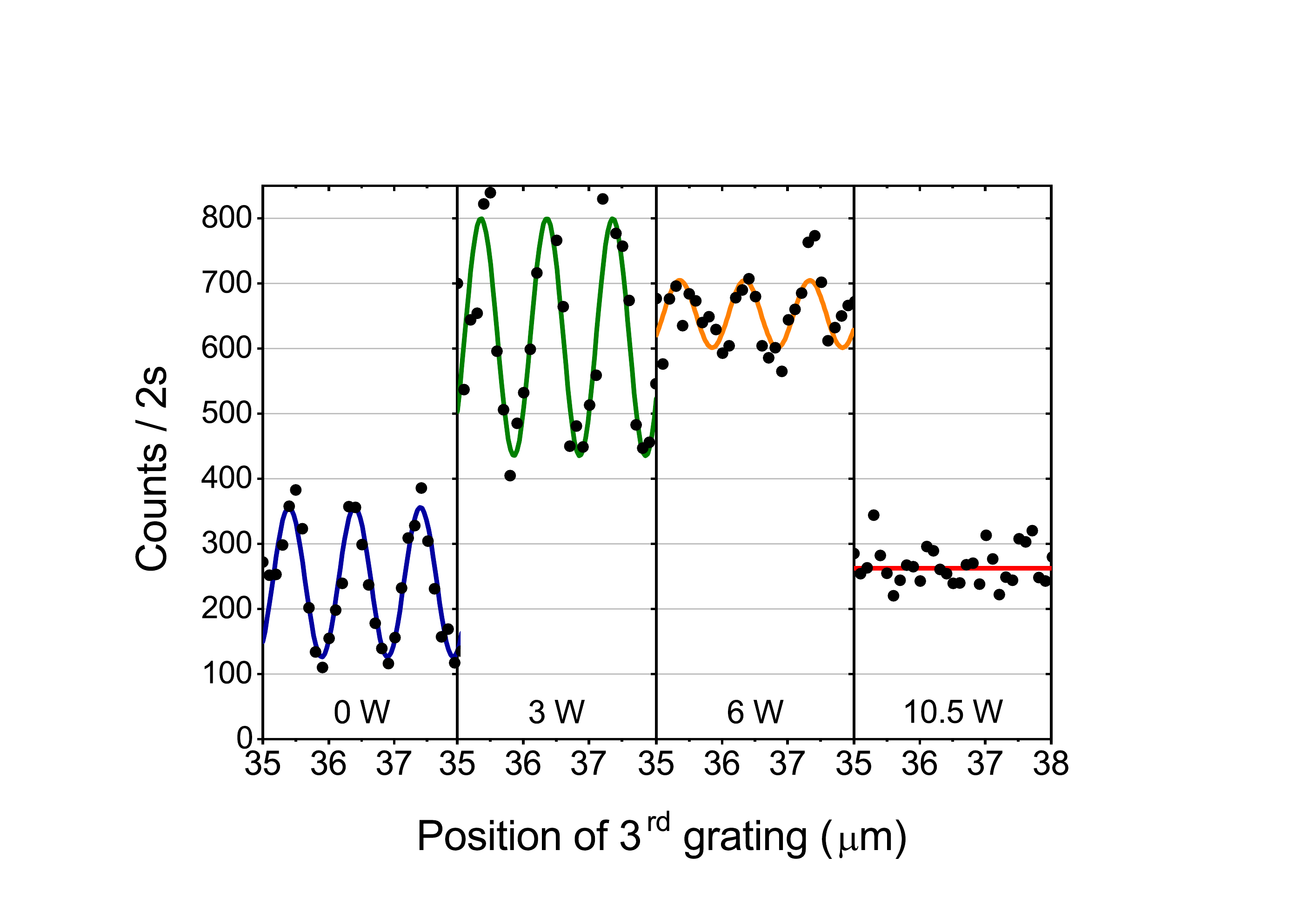}
    \caption{Interferograms of C$_{70}$ molecules in a TLI at different laser power settings in the heating stage. The count rate initially increases due to the increased thermal ionization probability at higher internal temperatures before it drops again at even higher laser power settings as a result of ionization and fragmentation in the heating stage. The interference contrast on the other hand drops continuously from initially 47\% to 0\% at $P=10.5$W due to the rising emission probability of position resolving photons within the interferometer. (Reproduced from Ref.~\cite{Hackermueller2004})
    \label{EDfig:Thermo_Interferograms}
    } 
\end{figure}

Decoherence due to thermal emission of radiation   was explored in an experiment on Talbot-Lau interference with C$_{70}$ fullerenes \cite{Hackermueller2004}. Multi-pass laser absorption was employed to heat the molecules internally to temperatures beyond 3000\,K, see Fig.~\ref{EDfig:Thermo_Interferograms}. 

\begin{figure}
    \centering
    \includegraphics[width=0.47\textwidth, trim=40 0 60 0,clip]{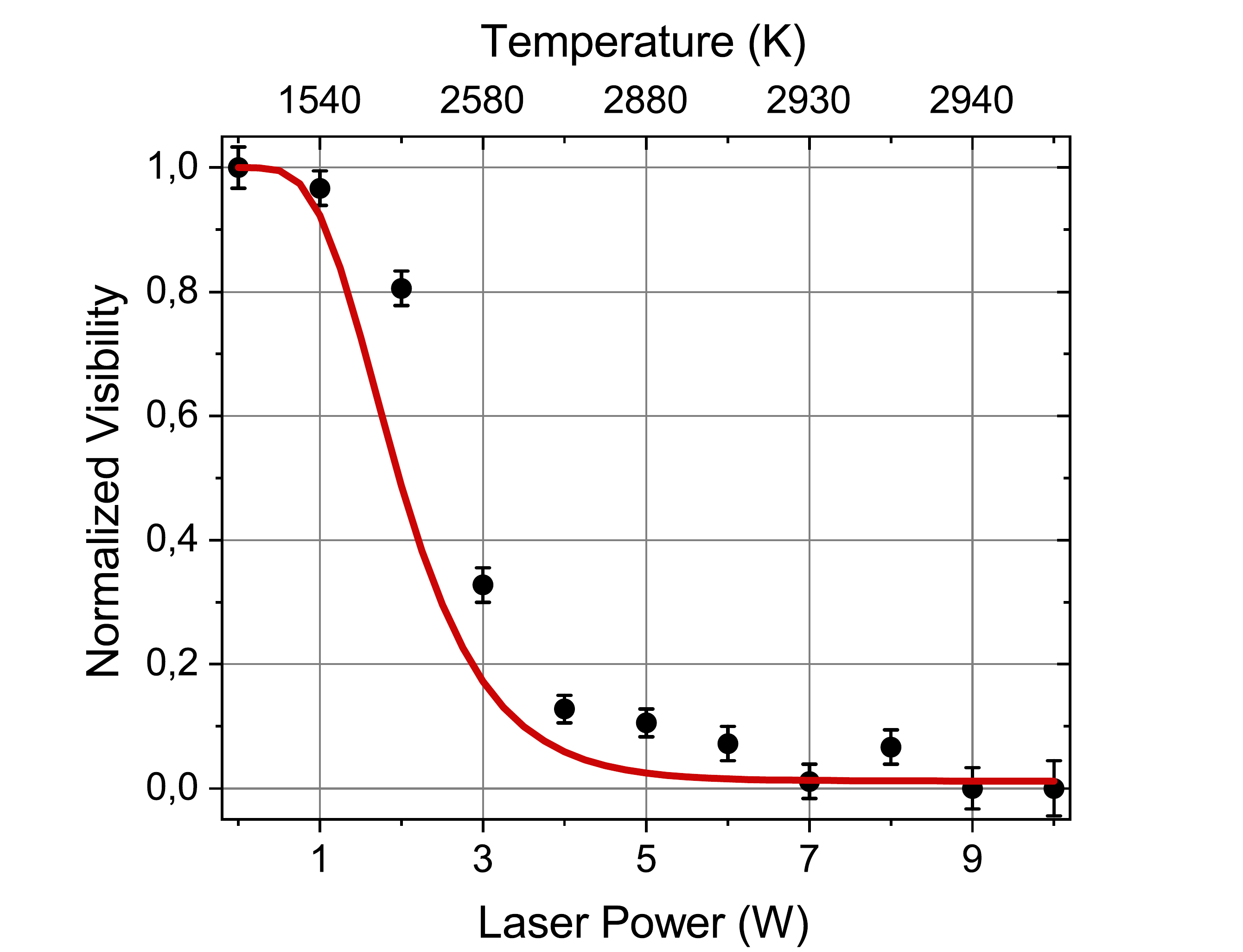}
    \caption{Loss of visibility due to thermal decoherence plotted versus heating power (lower scale) and mean internal temperature of the C$_{70}$ molecules when they enter the Talbot-Lau interferometer (upper scale) at a mean velocity of $v_m = 100$\,m/s. (Reproduced from Ref.~\cite{Hackermueller2004})\label{EDfig:Thermo_Visibility}}
\end{figure}

The maximum path separation in the interferometer exceeded 1$\,\mu$m so that even a single thermal photon with a wavelength of $2\,\mu$m emitted close to the second grating would reveal sufficient which-path information to fully decohere the interference pattern. A full theoretical 
description needs to account for the spectral distribution of the thermal radiation, the temperature decrease due to radiative cooling  \cite{Hackermueller2004,Hornberger2005}, and the emission position, since the  resolving power of a thermal photon is highest close to the middle grating and lowest close to the outer ones. Very good agreement was found between the predictions of decoherence theory and the experimental observation, see Fig.~\ref{EDfig:Thermo_Visibility}. It is striking that the internal temperature needs to exceed 1500\,K before sizeable decoherence could be observed, growing gradually and with the expected functional dependence when the temperature was increased. On a historical note, it is worth realizing that a source  temperature  of 900\,K was already required to obtain sufficient vapor pressure to perform the experiment. A small (and initially unforeseen) factor of two in temperature made the difference between the successful first realization of a molecule interferometer and the first observation of thermal decoherence.

Recent macromolecule experiments succeeded in demonstrating interference with up to 2000 atoms for a coherence time of about 10\,ms \cite{Fein2019}. While the emission of infrared photons is  to be expected in such a setting, the grating period  was a mere 266\,nm, about ten times shorter than the most probable photon emitted at 1000\,K. 
Since the photon wavelength is large compared to the path separation, momentum diffusion due to many weak photons is then the expected form of decoherence. This effect was still too weak to become influential but should be considered for even more complex particles.

\section{Perspectives in matter-wave decoherence }
\subsection{Experimental perspectives}
Interference experiments for matter waves are amongst the most sensitive devices for measuring forces, geometric phases, and accelerations. Atom interferometers  are about to reach path separations of half a meter \cite{asenbaum2017phase}, coherence times of half a minute \cite{xu2019probing}, and path lengths of hundreds of meters in free fall towers. 
Molecule interferometers, for which such length and time scales are still challenging, are venturing into the realm of biochemistry, targeting the delocalization of proteins, DNA and precursors of life, such as viroids and viruses. 
The quest of pushing matter-wave interferometry to ever larger objects was promoted by John~Clauser, who suggested delocalizing even \textit{little rocks and life viruses} \cite{Clauser1997}. The little rocks  are nowadays called nanoparticles, and viruses may not be considered genuinely alive, but an inspiration like this is at the heart of our own research program~\cite{Arndt1999,Arndt2005b,Juffmann2010}.

In a development complementary to studying freely falling matter wave beams, an increasing number of research groups aim at preparing \emph{trapped} nanoparticles in specific  quantum states of their center of mass. This requires cooling these objects into the deep quantum regime \cite{Chang2010} and will allow novel interferometric quantum experiments  
\cite{Romero-Isart2011,Arndt2014, Bateman2014a,Wan2016, Rudolph2020}.

Substantial progress in feedback cooling \cite{Li2011,Gieseler2012}, cavity cooling \cite{Asenbaum2013,Kiesel2013,Millen2015}, resolved side-band cooling \cite{Meyer2019}, and coherent scattering cooling \cite{Delic2019,Windey2019} brings many of these ideas closer to realization. The 1D ground state in a trap has been recently reached~\cite{Delic2020}. 

Beyond the center of mass, also the rotational motion of a nanoparticle may exhibit matter-wave interference. It gives rise to distinctive quantum recurrence phenomena \cite{Stickler2018,ma2020quantum}. Cooling and controlling the rotation dynamics of a levitated nanoparticle is therefore an important aim \cite{Hoang2016}.
In recent experiments our group  demonstrated precision optical control of silicon nanorotors \cite{Kuhn2017,Kuhn2017a}, suggesting that cooling to the rotational quantum regime will soon be feasible \cite{Stickler2016,Kuhn2017c,schafer2020cooling}. Matter wave interference in the rotational degrees of freedom will then be subject to orientational decoherence effects \cite{Stickler2016a,Papendell2017}.

\subsection{Probing new physics }
Aside from the many applications in sensing and metrology, there is another motivation for demonstrating matter wave interference with ever more massive and complex objects: The highly macroscopic quantum states \cite{Nimmrichter2013,Schrinski2020} involved allow probing unexplored territories where new physical effects might be found. While it is hard to say what to expect from the unknown, many suggestions and speculative ideas have been put forward that might be testable with matter-wave interferometry. 

For instance, it has been suggested that fluctuations of space-time on the Planck scale could have an observable effect \cite{Percival1997,Wang2006,Bonifacio2009}, even though  matter-wave experiments currently employ de Broglie wavelengths no less than $30$\,fm \cite{Fein2019}. The diffusion of these matter waves due to  space-time fluctuations might become detectable in the same way as agitated water molecules affect the motion of pollen in Brownian motion--provided one can boost the mass of the interfering objects by another six orders of magnitude. This sounds like a big step, and it is. But macromolecule interferometry operates already with masses almost eight orders of magnitudes greater than the electron, which stood at the beginning of matter-wave experiments.

It has also been suggested that decoherence due to gravitational waves might be observable \cite{Lamine2002,Lamine2006}. The random noise of gravitational waves would cause contractions of the arm length in a matter-wave interferometer and thus stochastically blur its fringes. A promising route will open up if one succeeds in creating large-momentum beam splitters for ultra-fast hydrogen atoms, for instance by diffracting 80\,eV hydrogen atoms coherently at the hexagonal 243\,pm lattice of single-layer graphene\cite{Brand2019}. 
As described in Sect.~\ref{EDsec:clocksingravitationalfield}, the interface between general relativity and  quantum physics can also be probed by delocalized atoms with a single fast-ticking internal clock or by molecules with slower but abundant vibrational clocks \cite{Zych2011,Pikovski2015}.

Decoherence could also be induced by particles beyond the standard model \cite{Riedel2017}.
In particular, numerous astronomical and cosmological observations hint at the existence of dark matter with an average density of 0.4\,amu/cm$^{3}$ in our galaxy. It might consist of  weakly interacting, massive particles (WIMPs) at low density or of ultra-light dark particles in quantum degenerate dark-matter waves (or of anything in between).
In case of high-mass WIMPs any of the rare collisions  with the matter-wave beam
would kick the nanoparticle beyond the detector area, so that dark matter leads to a loss of counts rather than decoherence. 
However, for the intermediate mass regime between 1\,eV/c$^2$ and $10^6$\,eV/c$^2$  collisions would be strong enough to cause decoherence while keeping the nanoparticles in the experiment. 
An observed additional loss of coherence would then be an indicator for dark matter--provided all natural decoherence processes are kept under control.
It has been suggested that this effect may become visible in matter-wave experiments with massive particles
if one can profit from a coherent enhancement of the scattering cross section
for  dark-matter de Broglie wavelengths greater than the size of the colliding partner \cite{Riedel2013,Riedel2015}.

Matter-wave interferometry is particularly well-suited to probe theories that postulate an objective, spontaneous collapse process for the quantum wave function \cite{Bassi2013a}. Such models replace the Schr\"odinger equation with a nonlinear stochastic equation, causing macroscopically delocalized wave functions to collapse randomly and thus effecting a transition from quantum to classical behavior. 
The most prominent one is the model of continuous spontaneous localization (CSL) devised by Ghirardi, Rimini, Weber, and Pearle  \cite{Ghirardi1986,Pearle1989}. It was introduced \emph{ad hoc}, but it can also be seen as a representative of a generic class of `classicalizing’ modifications of quantum mechanics derived from natural consistency requirements \cite{Nimmrichter2013}. Importantly, CSL also implies a weak gradual heating of macroscopic objects. 

There are two parameters in the CSL hypothesis, a localization length $r_c$ and a rate parameter $\lambda$. While there is no \textit{a priori} constraint for them, the model only serves its purpose of `solving' the measurement problem by reinstating objective realism on the macroscale, if the length $r_c$ is in the microscopic domain and the rate $\lambda$ is sufficiently large. Only then will macroscopic objects behave classically. On the other hand, experimental observations can provide upper bounds for $\lambda$ given $r_c$. Currently, the best bounds on the CSL parameters are from the observed lack of heating in classical setups \cite{piscicchia2017csl,Carlesso2017,Vinante2020}, but the predicted heating can be reduced by tweaking the model.

The most stringent constraints due to quantum experiments are set by atom interferometry for $r_c$ on the meter scale \cite{Kovachy2015} and by macromolecule interferometry for $r_c\simeq 100\,$nm \cite{Nimmrichter2011}.
Since the collapse rate $\lambda$ scales with the square of the interfering particle mass it is favorable to use large molecules. On the other hand, a special type of interference using large squeezed atomic Bose-Einstein condensates  might allow ruling out the entire relevant parameter space of CSL, provided the number of detected atoms can be resolved sufficiently well \cite{schrinski2020how}.

As the mass of the interfering particle is increased, one may wonder what happens when the superposed object becomes sufficiently massive to turn into a sizeable source of gravity. 
It has also been argued \cite{Penrose1996} that a massive object in superposition of position states will deform space-time locally and thus create a superposition of space-times. The latter might collapse on a time scale given by the energy-time uncertainty relation involving the gravitational self-energy.  
This can be cast into a dynamical decoherence-type equation \cite{Diosi1984} where the collapse rate  scales with the square of the mass, speaking again in favor of high mass interferometry. 

In a similar spirit, the idea of gravitational self-interaction may motivate the so-called  Schr\"odinger-Newton equation, leading to a rather unusual nonlinear evolution of the center-of-mass wavefunction, with a partial shrinking to a soliton-like solution and a partial spread to infinite distances \cite{Grossardt2013}.
Such a behavior would  be expected for particle masses of $10^{11}$\,Da only after several 10$^4$\,seconds of free evolution time, but the compatibility with special relativity 
is not clear  \cite{Giulini2018}.

Finally, recent studies have also suggested that one might probe the quantumness of gravity by placing two superposed micro-spheres in close vicinity: if their positions become entangled by gravity, gravity would need to be quantized. 
Such ideas will become testable if one succeeds in placing particles with masses in the range of $10^{14}\,$Da in superposition states with a delocalization range as wide  as $100\,\mu$m  \cite{Marletto2017,Bose2017}. 
Preparing and probing such a macroscopic superposition state is a highly ambitious goal and many decoherence studies will still be required to push matter-wave physics to this range.

The work by H.~Dieter Zeh and many of the pioneers of decoherence theory has paved the path for a better understanding of the limitations of modern quantum technologies and it has triggered a plethora of new experiments to improve our understanding of the foundations of physics. His name will always stay connected to decoherence research.
\acknowledgments
M.A. thanks Anton Zeilinger and Lucia Hackermüller, Bj\"orn Brezger and Stefan Uttenthaler for intriguing first years in molecule interferometry that culminated in our joint experiments on collisional and thermal decoherence as included and referenced here. We thank Filip Kialka for comments on the manuscript. 
We acknowledge funding by the European Commission within the project ERC AdG 320694 and the Austrian Science Funds in the project P-30176.

\end{document}